\documentclass{article}
\usepackage{graphicx, amsfonts, amsthm, amsmath, draftwatermark}
\newtheorem{definition}{Definition}[section]
\newtheorem{proposition}{Proposition}[section]

\SetWatermarkText{DRAFT}
\SetWatermarkScale{1}

\title{OpenAlpha: A Community-Led Adversarial Strategy Validation Mechanism for Decentralised Capital Management}
\author{Arman Abgaryan*, Utkarsh Sharma*}
\date{April 2025}

\newcommand\blfootnote[1]
{
  \begingroup
  \renewcommand\thefootnote{}\footnote{#1}
  \addtocounter{footnote}{-1}
  \endgroup
}

\begin{document}

\maketitle

\blfootnote{*Lead co-authors and designers.}

\begin{abstract}
    We propose \textit{OpenAlpha}, a community-led strategy validation framework for decentralised capital management on a host blockchain network, which integrates game-theoretic validation, adversarial auditing, and market-based belief aggregation. This work formulates treasury deployment as a capital optimisation problem under verification costs and strategic misreporting, and operationalises it through a decision waterfall that sequences intention declaration, strategy proposal, prediction-market validation, dispute resolution, and capital allocation. Each phase of this framework's validation process embeds economic incentives to align proposer, verifier, and auditor behaviour, producing confidence scores that may feed into a capital allocation rule. While OpenAlpha is designed for capital strategy assessment, its validation mechanisms are composable and extend naturally to evaluating external decentralised applications (DApps), enabling on-chain scrutiny of DApp performance, reliability, and integration risk. This architecture allows for adaptive, trust-minimised capital deployment without reliance on centralised governance or static audits.
\end{abstract}

\section{Introduction}
    Decentralised asset management is a critical yet underdeveloped component of the blockchain financial stack. While capital formation and trading have been extensively decentralised, treasury management, i.e. the allocation of pooled capital toward productive strategies, remains opaque and often centralised. In most blockchain ecosystems, protocol teams directly control treasury allocation despite lacking institutional-grade financial expertise, which introduces centralisation-based inefficiencies and governance risk, as capital may be deployed into low-conviction or misaligned strategies, or in worse cases, be susceptible to insider capture. More fundamentally, existing treasury governance frameworks, whether off-chain or on-chain, fail to scale under decentralisation. Many decentralised treasuries lack effective ways to source and evaluate strategies from the community. Their goals - such as profit maximisation, grant-based ecosystem growth, or liquidity incentives, often vary, making it difficult and costly to assess whether a proposed strategy is both suitable and reliable. This limitation runs counter to the foundational premise of blockchain: broadest possible permissionless access, decentralised decision making, without unfair barriers to entry. Even governance votes, as they're presently structured, cannot by themselves resolve whether a strategy has over-fit historical data, misrepresented liquidity assumptions, or encoded hidden risk. As a result, decentralised treasuries are either under-deployed or allocated in ways that do not reflect the broader community’s risk and return preferences.\\
    \\
    We formalise decentralised capital management as a principal-agent allocation game under asymmetric information and verification costs, framed as an adversarial, partially observable capital allocation problem (i.e. strategy proposers have private alpha, which is costly to verify) subject to strategic misreporting. Put simply: \textit{how do we let anyone propose lucrative strategies, but only fund the good ones}? We introduce a framework where strategy proposers submit candidate strategies, which triggers a decentralised verification process conducted by community members who participate in binary prediction markets, deep audits, and challenge resolution games. Thereafter, the system aggregates market-based confidence scores with realised performance to continuously re-weight capital allocations.\\
    \\
    The proposed mechanism is a generalisable architecture, which can also be applied to adjacent domains, for e.g., DApp risk scoring, on-chain competitive learning of collective objectives, and a framework for incentive-compatible open-source collaboration. In this extended setting, the framework may treat not only strategies, but also the DApps with which those strategies interact as validation targets, such that each DApp may be scored on dimensions such as execution reliability capital utilisation efficiency, and technical and economic security, using the same market-driven validation and dispute resolution logic. Thereby paving the way to ensure that capital is routed not only to effective strategies, but to the most reliable and systemically safe DApps. This enables the formation of a dynamic, bottom-up capital flow regime where both strategic logic and downstream protocol integrations are adversarially validated, scored, and weighted in allocation. In effect, OpenAlpha becomes a decentralised \textit{affirmation} layer for risk-aware, trust-minimised capital deployment across the DeFi ecosystem.\\
    \\
    At the heart of our framework is a decision waterfall, a phased mechanism that sequences capital intention declaration, strategy submission, prediction-market-based validation, resolution (by deep searchers and committees), and final allocation. Each phase is incentive-aligned and formally contributes to a Bayesian update over the strategy’s validity (so on-chain markets dynamically learn which strategies truly work). This architecture replaces static governance with a dynamic, adversarial evaluation process grounded in information theory.\\
    \\
    Key innovations incorporated in our work include:
    
    \begin{itemize}
        \item A \textit{crowd-sourced, incentive-compatible capital optimisation model}, which maximises the utility of deployed strategies for a decentralised capital owner. By internalising verification costs (see Theorem XXX) as functions of community participation, this model enables scalable, adversarially robust capital deployment—without relying on centralised oversight or fixed auditing budgets.
        \item An \textit{intention-specification phase}, allowing capital owners to express parametric goals over strategy metrics (e.g., maximise return, minimise volatility). This provides a clear, programmable benchmark for evaluating proposals, ensuring capital is only allocated to strategies aligned with transparent, user-defined objectives.
        \item A \textit{permissionless strategy proposal mechanism}, where agents (for e.g., AI agents) submit candidate strategies. The proposed mechanism creates an open marketplace for capital management ideas—lowering the barrier to entry for new strategies while filtering out low-quality proposals via economic commitment.
        \item A \textit{prediction-market-based validation protocol}, in which community verifiers vote on binary claims of alignment between a strategy and its declared intention, producing a market-implied confidence score, which is eventually settled with aggregated expertise.
        \item A \textit{multi-stage dispute resolution mechanism} incorporating (i) challenge voting by reputation\footnote{Reputation is defined in terms of the track record of an agent having performed validation tasks, gleaned from their holdings of Alpha Tokens, representing reputation in OpenAlpha.}-weighted deep searchers, and (ii) fallback committees for critical disagreements. This ensures that even complex, high-risk proposals can be evaluated with rigour—preserving protocol safety.
        \item An \textit{stochastic audit lottery}, modelled as a Poisson process, that enforces probabilistic spot checks on dormant or unchallenged strategies. This guarantees a baseline adversarial pressure on all strategies over time, increasing the cost of undetected fraud while keeping verification costs efficient and non-deterministic.
        \item A \textit{modular, composable scoring system} in which capital is sized and allocated through a meta-allocation function over nested intentions, formally treated may be treated as a Bayesian inference problem. Each intention class serves as a structured prior over expected outcomes, with posterior updates driven by both strategy-level performance, which reframes capital deployment as belief-weighted bet sizing across intention-conditioned subspaces - enabling the system to adaptively reinforce not just high-yielding strategies, but the predictive validity of broader strategic goals and their validators.
        \item An \textit{incentivised validation mechanism} for strategy proposers and target marketplaces to fund decentralised verification (through OpenAlpha), as a pre-requisite for access to allocator capital and associated managerial fees.
        \item A \textit{filtering process} that allocates limited assessment resources to high-probability strategies, enhancing capital deployment efficiency.        
    \end{itemize}

    \noindent
    In summary, we propose a decentralised verification and capital allocation framework that reinterprets strategy evaluation as a continuous, incentive-driven market process, rather than a static governance vote. By integrating adversarial validation, prediction-market incentives, and adaptive Bayesian feedback into a unified architecture, OpenAlpha enables trust-minimised, economically robust, and dynamically reconfigurable capital deployment - laying the foundation for scalable, decentralised financial coordination across on-chain ecosystems.

\section{Problem Formulation}
    We model decentralised capital management as a sequence of capital‐allocation decisions made at discrete epochs $t=0,1,2,\dots$. At each epoch, the protocol must distribute a fixed capital budget $C>0$ across a pool of candidate strategies, trading off expected returns against on‐chain verification costs and strategic misreporting.

    \begin{definition}[Strategy Proposal Set]
        Let $\mathcal{S}_t = \{S^1, S^2, \dots, S^{n_t}\}$ denote the set of $n_t$ active strategy proposals at epoch $t$, where each $S^i$ is an autonomous strategy logic deployed on‐chain.
    \end{definition}

    \begin{definition}[Capital Intentions]
        A capital owner $a$ declares an \emph{intention set} - $\mathcal{I}_a = (Q_a, R_a, D_a)$, which is comprised of:
        \begin{itemize}
            \item Decision criteria: $Q_a = \{q_1, q_2, \dots, q_m\}$, where each $q_j: \mathcal{S} \to \{0,1\}$ is a binary-valued predicate representing a decision rule specified by the capital owner, such that each predicate encodes a binary condition, for example: $q^j(S^i) = \mathbb{1}[R_{S^i} < x\%]$, indicating whether the strategy $S^i$ satisfies (or violates) a threshold-based constraint. The set $Q_a$ enables declarative filtering of candidate strategies prior to or in parallel with prediction market validation.
            \item $R_a \in \mathcal{T}$, the readjustment frequency, defines the desired cadence for strategy re-evaluation and capital reallocation. $\mathcal{T} \subset \mathbb{N}$ is a discrete set of allowable epochs (e.g., daily, weekly, monthly), determining how often the intention set is re-applied to update allocations.
            \item Validation Criteria - $D_a \in \mathcal{D}$, is a decision override criteria which intention specifier uses to state whether they would like to go with the crowd sourced decision, for e.g., vote-majority thresholds (e.g., $\text{Agree} \geq 70\%$), divergence tolerances (e.g., $\sigma(\omega^i) \leq \delta$), role-based triggers (e.g., requiring validation by a deep searcher or final arbitration), etc.
        \end{itemize}
        
        \noindent
        Collectively, the tuple $\mathcal{I}_a$ encodes a modular, programmable specification of the capital owner's preferences, enabling the system to filter, validate, and allocate strategies in a decentralised and formally interpretable manner.
    \end{definition}
    
    \begin{definition}[Strategy Utility]
        Let a capital allocator $a \in \mathcal{A}$ assign an amount $A_a^i(t) \geq 0$ to strategy $S^i$ at epoch $t$, using the realised contribution of a strategy, which is captured by a value vector: 
        \begin{equation}
            \Delta \mathbf{V}^i(t) := M_a(S^i, A_a^i(t)) \in \mathbb{R}^k,
        \end{equation}

        \noindent
        where $M_a: \mathcal{S} \times \mathbb{R}_+ \to \mathbb{R}^k$ is the allocator’s multi-objective metric function (as defined in the intention tuple $\mathcal{I}_a$), mapping strategy-performance outcomes to capital-relevant evaluation dimensions such as return, risk, liquidity impact, or volatility.\\
        \\
        The allocator assesses the realised contribution using a utility function: $U_a^i: \mathbb{R}^k \to \mathbb{R}$ denote the allocator’s utility function, which encodes their preference over the $k$-dimensional metric space (e.g., return, volatility, drawdown, and liquidity impact)\footnote{Whilst the specific formulation of this utility function is left for future work, $U_a^i$ can be assumed to be monotonic in beneficial directions and concave to reflect diminishing marginal utility or risk aversion.}. This function governs how allocator $a$ derives subjective benefit from the capital deployed to $S^i$, guiding their future allocation updates.
    \end{definition}
    
    \begin{definition}[Verification Cost Function]
        Let $\Phi^i\big(A^i(t)\big) \in \mathbb{R}_{\geq 0}$ denote the total cost incurred by the protocol in verifying the correctness, robustness, and safety of strategy $S^i$ given capital allocation $A^i(t)$ at epoch $t$.\\
        \\  
        This cost may include (but is not limited to): (i) incentives paid to verifiers and auditors (e.g., prediction market rewards, deep searcher bounties); (ii) computational or gas costs associated with validation infrastructure; (iii) expected expenditure from audit lotteries or fallback dispute mechanisms, etc.\\
        \\
        The function $\Phi^i(\cdot)$ is assumed to be non-decreasing in $A^i(t)$, and may also depend on strategy-specific complexity characteristics (e.g., number of contracts, depth of historical backtests). But in general, $\Phi^i$ is endogenous to system behaviour: community participation, adversarial response probability, and prediction market liquidity all influence realised verification costs. A fully dynamic model of $\Phi^i$ would require equilibrium analysis of validator incentives and search effort, which we leave to future work.
    \end{definition}
    
    \noindent
    The manager’s objective is to allocate capital $C \in \mathbb{R}_{>0}$ across a set of candidate strategies $\mathcal{S}_t = \{S^1, \dots, S^{n_t}\}$, so as to maximise the aggregate expected utility derived by all capital allocators, net of verification costs:
    
    \begin{equation}
        \begin{aligned}
            \max_{\{A^i(t)\}_{i=1}^{n_t}} \quad & \sum_{i=1}^{n_t} \left[ \mathbb{E}_{r^i(t) \mid \omega^i(t)}\left[ U_a^i\big(\Delta V^i(t)\big) \right] - \Phi^i\big(A^i(t)\big) \right] \\
            \text{s.t.} \quad & \sum_{i=1}^{n_t} A^i(t) = C, \quad A^i(t) \geq 0 \quad \forall i.
        \end{aligned}
    \end{equation}

    \noindent
    Where $\Delta V^i(t) := r^i(t) \cdot A^i(t)$ is the realised value created by strategy $S^i$; $U_a^i: \mathbb{R}_{\geq 0} \rightarrow \mathbb{R}$ is the utility function specific to the capital allocator $a$; $\omega^i(t) \in [0,1]$ is the market-derived belief signal (confidence score); $\mathbf{r}^i(t) \in \mathbb{R}^k$ denote the random vector of realised performance outcomes of strategy $S^i$ at epoch $t$, evaluated over $k$ distinct metrics (e.g., return, volatility, drawdown, liquidity), such that conditional on the information set $\mathcal{F}_t$ (which implicitly includes past outcomes, validation signals, and other relevant data up to epoch $t$), we can model $\mathbf{r}^i(t)$ as being drawn from a posterior distribution: $r^i(t) \sim \mathcal{P}(r^i \mid \omega^i(t), \mathcal{F}_t)$; and $\Phi^i(A^i(t))$ denotes the total endogenous cost of verifying $S^i$, which may also depend on system-wide community behaviour, strategy complexity, or market incentives.\\
    \\
    In practice, at each rebalancing epoch the static optimisation is executed through our decision waterfall:
    
    \begin{enumerate}  
        \item \textit{Intention Phase}: Capital owners declare $\{\mathcal I_a\}$, fixing the metric–goal pairs that parameterise $U^i $ and shape the verification criteria.  
        \item \textit{Proposal Phase}: Agents submit strategies $S^i $ and post collateral, populating  $\mathcal S_t $ and linking each  $S^i $ to a subset of intentions.  
        \item \textit{Validation Phase}: Community members open LMSR markets on binary claims $\{C_{i,j}\}$, generating prices $p_{i,j}(t)$, and thus confidence scores $\omega^i(t) $.  
        \item \textit{Resolution Phase}: Deep searchers and review committees stake reputation (Alpha tokens) and collateral to resolve disputes, finalising which claims count toward  $\omega^i(t) $ and allocating the corresponding $\Phi^i_{\rm pm}$ and $\Phi^i_{\rm ds} $ costs.  
        \item \textit{Allocation Phase}: A capital allocator may compute $A^i(t)$ by solving the constrained optimisation above (using $\omega^i(t) $ to adjust the expected utility terms and charging verification fees $\Phi^i$), and rebalances the capital allocation accordingly.
    \end{enumerate}

    \noindent
    Over multiple epochs, this is effectively a problem of the dynamic programming type,
    \begin{equation}
         \max_{\{A^i(t)\}} \mathbb{E}\Bigl[\sum_{t=0}^T \delta^t \bigl(U(\Delta V(t)) - \Phi(A(t))\bigr)\Bigr],
    \end{equation}

    \noindent
    where $\delta\in(0,1)$ is a discount factor, and both returns $r^i(t)$ and claim‐market evolution $p_{i,j}(t)$ can be modeled to follow stochastic processes. An off-chain solver can implement this in a rolling-horizon fashion.
    
\section{Architecture}
    The operational framework revolves around distinct roles, performed by agents\footnote{These agents can either be an AI agent, or managed by a human.}, and incorporated in the system architecture through a fully auditable rules-based framework. Before we define the logic of the architecture, we define a few agents, and then define the decision-sequence.
    
    \begin{itemize}
        \item \textit{Proposer}: This agent is responsible for submitting a candidate strategy and initiating the validation process by formally linking the strategy to a declared capital intention. This linkage constitutes a public assertion, backed by economic commitment, that the proposed strategy is designed to fulfil the objectives specified by the intention. To deter frivolous or misaligned submissions, the proposer must stake a threshold number of affirmative (\lq\lq Agree\rq\rq) votes, representing a form of upfront economic signal. If the strategy passes validation and delivers verifiable net value to the capital allocator, the proposer is eligible to receive a performance-linked commission, typically parameterised as a negotiable fraction of realised returns.
        
        \item \textit{Verifier}: This agent participates in the decentralised validation layer, engaging in prediction markets and stake-weighted voting to assess the alignment of proposed strategies with declared intentions. Verifiers earn rewards based on the accuracy of their assessments, drawn from a market-funded pool and supplemented by protocol-issued reputation tokens. Incorrect votes (i.e., those misaligned with the eventual resolution) result in forfeiture of staked capital, which is redistributed to accurate participants, maintaining incentive alignment and discouraging opportunistic signalling.
        
        \item \textit{Deep Searcher}: Functioning as a second-layer validator, this agent conducts higher-order review (relative to community verifiers) of the prediction market outcomes, either challenging or affirming the preliminary consensus. Deep searchers are activated when the consensus confidence is low (or when discrepancies are flagged\footnote{How?}). They stake reputation tokens and are rewarded contingent on the final validation outcome. Accurate decisions enhance their influence and access to future high-tier validations, while repeatedly overturned decisions lead to slashing or demotion within the protocol’s trust framework.

        \item \textit{Arbitrators}: In the event of unresolved disputes (particularly where deep searcher outcomes are themselves contested) arbitrators conduct an extensive audit process, Arbitrators - next level of agents, operating with enforced neutrality and rigorous qualification requirements, perform independent reviews of both the strategy and the validation history. Final adjudication is based on protocol-defined criteria, and arbitrators are rewarded or penalised according to the consistency of their rulings with eventual outcomes and community consensus over time.
    \end{itemize}
    
    \noindent
    The system's core validation architecture can be visualised using the schematic, which is as follows:
    
    \begin{figure}
        \begin{center}
            \includegraphics[scale=0.4]{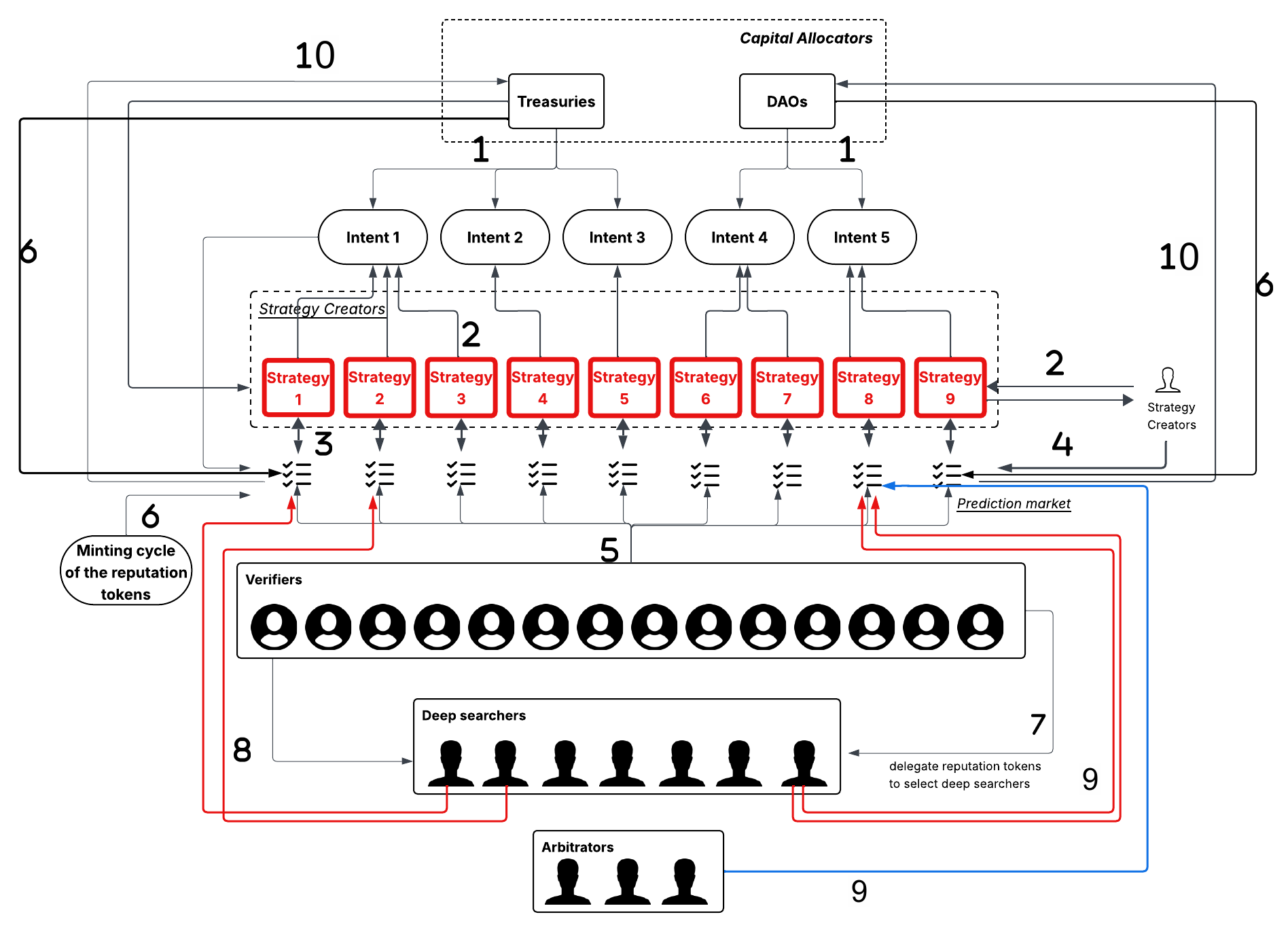}
            \label{fig:blocks}
            \caption{Summary schematics.}
        \end{center}
    \end{figure}  

    \noindent
    This schematic can be understood using the decision waterfall, which is the phased elucidation of the system's architecture. The decision waterfall is comprised of a sequence of interlinked phases: (a) an intention specification phase, (b) a strategy proposal phase, (c) a validation phase, and (d) a proposal resolution phase.
    
    \begin{enumerate}
        \item \textit{Intention Phase}: A capital allocator $a \in \mathcal{A}$ declares an intention tuple:
        \begin{equation}
            \mathcal{I}_a = (M_a, G_a, R_a, D_a), \quad M_a \in \mathcal{M}, \quad G_a \in \mathcal{G}, \quad D_a \in \mathcal{D}
        \end{equation}

        \noindent
        which encodes $M_a: \mathcal{S} \to \mathbb{R}$; incorporating the goal $G_a$ (e.g., $\min$, $\max$, etc.) - forming a parametric preference over outcomes; $R_a: \mathbb{R}^+$ - indicating the number of epochs after which the intention specifier would like to enable to intent to be re-opened for validation (to spot if a better strategy has emerged since initial allocation); and $D_a: ...$ - representing the decision criteria for the intent, i.e. whether they'd be satisfied with a simple majority of the community verifier vote, or require a specific threshold to be met, to avoid mandating deep searchers and arbitrators to settle the market.\\
        \\
        Note, that the intention specifier must deposit an initial investment amount of capital before the validation phase can be opened, proving availability of a baseline capital, and eventually is the party which pays for crowd assessment (burning a specific amount of Alpha tokens).
        
        \item \textit{Strategy Proposal Phase ($\mathcal{S}$)}: A strategy proposer $p \in \mathcal{P}$ submits a strategy: $\sigma_p: \mathcal{E} \rightarrow \mathcal{S}$, which maps environment variables ($\mathcal{E}$) to system states ($\mathcal{S}$), such that the strategy achieves the goals of a particular intention ($\mathcal{I}_a$), and only becomes activated when the decision criteria ($D_a$) of the intention specfier has been met - whether it's receiving enough \lq\lq agree\rq\rq votes in the prediction market, or market mandatorily being settled by the deep searchers.\\
        \\
        Note, that the strategy space evolves both prior to and following the declaration of capital owners’ intentions. This reflects an endogenous interdependence, i.e. a \lq\lq chicken and egg\rq\rq problem: attracting high-quality strategies requires clearly specified intentions, while incentivising capital owners to declare intentions benefits from the presence of credible, high-performing strategies. As such, the process is inherently dynamic and mutually reinforcing, and presence of strategies which do not immediately serve an intention may serve as a mechanism for continuously attracting additional capital commitments into the system. And since we do not charge intention specifiers until their strategy is deployed, it is also possible for capital owners to declare their intentions, and await emergence of a strategy which helps them meet their pre-defined goals.
            
        \item \textit{Validation Phase ($\mathcal{V}$)}: The validation phase consists of six substeps:
            \begin{enumerate}
                \item \textit{Initiation}: A proposer $p$ links $\sigma_p$ with one or more intentions $\mathcal{I}_a$, creating a validation instance:
                \begin{equation}
                    \mathcal{V}_{p,a} = (\sigma_p, \mathcal{I}_a),
                \end{equation}

                \item (Validation) Prediction Market Inception: The proposer stakes/deposits an amount ($c_p > 0$ tokens) to initiate the binary prediction market $\mathcal{M}_{p,a}$, which counts towards initial set of positive/agree votes. As such, the market starts with:

                \begin{equation}
                    T_0^{\text{Agree}} = c_p
                \end{equation}

                \noindent
                which represents the initial amount staked for the agree side.

                \item \textit{Voting}: Each community verifier ($v \in \mathcal{V}$) commits $t_v$ tokens to either side of the market:
                \begin{equation}
                    \text{Vote}_v \in \{\text{agree}, \text{disagree}\}, \quad t_v \in \mathbb{R}_{>0},
                \end{equation}

                \noindent
                such that an \lq\lq agree\rq\rq vote implies that the proposed strategy meets a particular intention specifier's goals, and \lq\lq disagree\rq\rq implies that it does not. At the end of the voting period, we would then have the total votes:
                \begin{equation}
                    T_t^{\text{agree}}, \quad T_t^{\text{disagree}}.
                \end{equation}

                \noindent
                Here, incremental tokens required for each votes are derived using a Parimutuel mechanism \cite{koessler2008parimutuel}, and over time, for a select number of premier strategies - we may use the LMSR mechanism \cite{hanson2003combinatorial}, reasons for which are discussed in forthcoming sections.

                \item \textit{Resolution}: Unless, the intention specifier indicates they're happy to progress with the decision of the community, the market resolves once the decision criteria of the intention specifier has been met (sufficient \lq\lq agree\rq\rq votes, or through settlement by deep searchers $d \in \mathcal{D}$, who stake a threshold amount of Alpha tokens $r_d \geq r_{\min}$).
                
                \item \textit{Arbitration Window}: After the market is resolved (either by a deciding vote of deep searchers or a supermajority of community votes) an arbitration window is opened. During this period, an arms-length arbitrator\footnote{Arbitrators must meet baseline qualification thresholds, such as participation in a minimum number of prior disputes. Their rewards are denominated in a reputation-weighted token with meaningful value, ensuring credible incentives.} may challenge the resolution outcome by staking a required amount of Alpha tokens. Crucially, this challenge is financially at risk: if the arbitrator’s challenge is upheld, the original outcome $o_d$ is reversed ($o_d \mapsto \neg o_d$) and the reputation stake of the originally deciding deep searcher is slashed ($r_d \to 0$). If the challenge fails, the arbitrator’s own staked reputation is forfeited. This creates a credible deterrent against both collusion and frivolous escalation.
                
                \item Settlement: Winners earn the entire staked amount of losing side's token, and additionally become eligible to receive the protocol's Alpha tokens, i.e. based on an eligibility function $\phi: \mathcal{V} \to {0,1}$.
            \end{enumerate}
            
        \item \textit{Allocation Phase ($\mathcal{A}$)}:  In the final stage of the decision waterfall, an allocator may use the OpenAlpha's outcomes to allocate their capital across validated strategies according to an intention-specific meta-allocation function $f_a(\omega^i, \xi^i)$, where each $f_a$ governs how the confidence score $\omega^i$ (derived from prediction markets), relative amount of \lq\lq agree\rq\rq votes, and environmental variable $\xi^i$ (capturing dynamic context, such as asset volatility or liquidity depth) are weighted under capital owner $a$’s declared objective. Quite crucially, the inclusion of vote-weighting in the allocation rule incentivises proposers to compete not only on strategy quality but also on generating broader community consensus, as more \lq\lq agree\rq\rq votes increase the likelihood of a strategy receiving capital. Furthermore, the space of such allocation functions may itself be subject to competitive refinement: agents may propose updated formulations of $f_a$ or re-weighting of intention class priors, for e.g. - represented as Dirichlet hyperparameters $\alpha_a$, and have them validated via a protocol-level challenge-and-endorsement process. This allows the system to meta-learn over time not only which strategies perform best, but which goal functions and allocation heuristics most consistently generate high-confidence, high-yield deployments under varying market regimes. The result is  hierarchically Bayesian, self-improving allocation mechanism that fuses validated strategy outcomes, community trust signals, and real-time market conditions within a modular and decentralised framework.
    \end{enumerate}

    \noindent
    The aforementioned decision waterfall translates into OpenAlpha protocol having a lifecycle, which is structured across distinct, non-overlapping time intervals, which regulate how strategies are proposed, assessed, executed, and exited. If we use $\tau \in [0, \infty)$ to represent a time interval of a specific phase, then we have: 
    
    \begin{itemize}
        \item Strategy Proposal Interval: This is the interval during which the strategy proposers can formally submit candidate strategies ($\sigma \in \Sigma$) for consideration for specific intents, which must be accompanied by a minimum amount of tokens ($c_\text{min}$)/\lq\lq agree\rq\rq votes to avoid non-frivolous assertion that a strategy meets an intention specifier's requirements.
        
        \item Assessment Interval: During this phase, the community evaluates submitted strategies using decentralised prediction markets, and verifiers vote on whether each strategy is aligned with the declared intention set $\mathcal{I}$ of capital allocators. Challenges to validity are raised here, deep searchers resolve conflicting votes by staking Alpha tokens, and arbitrators resolve disputes between deep searchers.  The output of this phase is a validated set of executable strategies.
        
        \item Rebalancing Interval: Strategies validated in $\tau_a$ and allocated capital, are available for re-allocation in this phase, when based on the requirements of the capital allocator - the  available strategies can be opened for reassessment, opening the possibility for rebalancing capital between different strategies.
        
        \item Withdrawal Interval: This is the exit window during which capital allocators and strategy executors can withdraw proceeds, including profits or residual funds from the strategy. Here, realised performance metrics are collected and evaluated against the original intention set $\mathcal{I}$, potentially feeding back into future reputation or scoring mechanisms.  
    \end{itemize}

\section{Prediction Market for Strategy Validation}
    To operationalise decentralised verification in a fully permissionless setting, we embed a family of binary prediction markets into the strategy evaluation pipeline. These markets transform the verification process into a continuous-time, market-driven scoring game, wherein community participants stake probabilistic beliefs on the alignment between a submitted strategy and a predefined capital objective. Rather than atomising individual verification tasks or requiring centralised expert audits (unless absolutely necessary in final stages of the validation process), this design abstracts risk assessment into an incentive-compatible signalling process. The resulting market prices serve as belief-weighted estimators of strategic alignment, enabling the system to aggregate distributed, adversarial scrutiny into a unified, endogenous confidence metric - that can be gleaned from the final odds.\\
    \\
    For each strategy $S^i$ and each intention $\mathcal I_a=(M_a,G_a)$, define the binary claim:
    \begin{equation}
        C_{i,a} := \bigl \lbrace G_a\bigl(M_a(S^i)\bigr) \bigr\rbrace,
    \end{equation}

    \noindent
    which is intended to test the assertion that $S^i$ achieves the goal $G_a$ on metric $M_a$. This claim $C_{i,a}$ is initiated when a strategy proposer stakes a slashable amount/bond ($b_{i,a}>0$), which funds the initial liquidity - \lq\lq agree\rq\rq votes.\\
    \\
    In our primary implementation, we adopt the Parimutuel mechanism \cite{koessler2008parimutuel} as the core structure for strategy validation markets. Unlike scoring-rule-based mechanisms (e.g., LMSR), Parimutuel markets require no budgeted market maker and impose no exogenous loss on any entity - making it quite suitable for the initial rollout. Here, participants bet directly against one another: total stakes on the losing outcome are redistributed to winners pro rata, based on their relative contribution to the winning side. Furthermore, under mild assumptions of risk neutrality and independent beliefs, it can be shown that implied odds converge in expectation to the true event probability, as shown in equilibrium analyses such as \cite{koessler2008parimutuel}. However, Parimutuel may underperform in cases where there is thin participation, leading to situation where early trades can strongly skew implied odds and discourage informed entry. Therefore, as the protocol matures and validator participation increases, transitioning to a Logarithmic Market Scoring Rule (LMSR) can provide continuous pricing, bounded loss guarantees, and smoother information aggregation. LMSR introduces a cost function that incentivises marginal information revelation even in low-liquidity settings, which makes it particularly effective for maintaining signal fidelity in long-tail claims or under sparse verification activity, whilst also providing arbitrage-free sequential updates, and full control using the liquidity parameter (which can be dealt via some governance).\\
    \\
    Now, in the Parimutuel setting, the claim  $C_{i,a}$ operates as a binary staking pool where each participant $v$ selects a binary position $o_v \in \{\text{Yes}, \text{No}\}$ and commits stake $s_v > 0$. At the end of the betting phase, stakes are aggregated into $\mathcal{P}_{o^*}^{i,a}$ - the total stake on the winning outcome and $\mathcal{P}_{\neg o^*}^{i,a}$ - the total stake on the losing outcome. Once deep searchers resolve the market, the winning side is paid proportionally from the losing pool:
    \begin{equation}
        \text{Payout}_v =
            \begin{cases}            
                s_v \cdot \dfrac{\mathcal{P}_{\neg o^*}^{i,a}}{\mathcal{P}_{o^*}^{i,a}}, & \forall o_v = o^*,\\
                0, & \forall o_v \neq o^*,
            \end{cases}
    \end{equation}

    \noindent
    where $o^*$ is the resolved outocme of the validation market. This mechanism yields an implicit belief estimate for claim $C_{i,a}$:

    \begin{equation}
        \hat{p}^{i,a}_{\mathrm{Y}} = \frac{\mathcal{P}_{\mathrm{Y}}^{i,a}}{\mathcal{P}_{\mathrm{Y}}^{i,a} + \mathcal{P}_{\mathrm{N}}^{i,a}},
    \end{equation}

    \noindent
    which is interpretable as the crowd-implied probability that $S^i$ satisfies the capital owner’s goal $G_a$ over metric $M_a$. This quantity can be used analogously to the LMSR price in computing confidence-weighted allocation scores.\\
    \\
    Now, to ensure continuous liquidity and a bounded‐loss market, we may employ a Logarithmic Market Scoring Rule (LMSR) with liquidity parameter $\ell>0$.

    \begin{definition}[LMSR Cost Function]
        Let $q^{i,a}_{\mathrm{Y}}$ and $q^{i,a}_{\mathrm{N}}$ be the cumulative \lq\lq yes\rq\rq and \lq\lq no\rq\rq shares for claim $C_{i,a}$. The cost to purchase an additional $\Delta q$ yes‐shares is:
        
        \begin{equation}
            \Delta C = \ell\ln\bigl(e^{(q_{\mathrm{Y}}^{i,a}+\Delta q)/\ell}+e^{q_{\mathrm{N}}^{i,a}/\ell}\bigr) - \ell\ln\bigl(e^{q_{\mathrm{Y}}^{i,a}/\ell}+e^{q_{\mathrm{N}}^{i,a}/\ell}\bigr).
        \end{equation}

        \noindent
        The instantaneous market price (belief) of \lq\lq yes\rq\rq is:
        \begin{equation}
            p^{i,a}_{\mathrm{Y}} = \frac{e^{q_{\mathrm{Y}}^{i,a}/\ell}}{e^{q_{\mathrm{Y}}^{i,a}/\ell}+e^{q_{\mathrm{N}}^{i,a}/\ell}}.
        \end{equation}

        \noindent
        This means that the worst case loss the market maker of this binary market is capped by $\ell\ln2$, which in turn allows us to backward work the value of the liquidity parameter $\ell$. This bounded subsidy is financed by payment from the strategy proposer's staked amount, which can be topped up by Reputation Token owners (if need be) - who may be incentivised to facilitate a liquid and smooth prediction market.
    \end{definition}  

\section{Audit Lottery}
    To guarantee a minimum level of adversarial scrutiny on every strategy (independent of community-raised claims), we embed a stochastic audit lottery into the decision waterfall, which acts as a baseline verification layer, complementing prediction‐market validation and adversarial validation.

    \begin{definition}[Audit Intensity]
        For each active strategy $S^i\in\mathcal S_t$, let $\Delta t$ denote the elapsed time since its last formal verification.  We model audits as a Poisson process of rate $\lambda>0$, so that the probability of at least one audit occurring in $\Delta t$ is:
        \begin{equation}
            \pi^i_t = 1 - \exp(-\lambda \Delta t).
        \end{equation}
            
    \end{definition}
    
    \noindent
    By construction, $\pi^i_t$ grows toward 1 for long‐untested strategies, ensuring eventual audit probability will increase over time.
    
    \begin{proposition}
        Over a horizon of length $T$, the number of audit events for $S^i$ is Poisson$(\lambda\,T)$.  In particular,
        \begin{equation}
            N^i(T)\sim\mathrm{Poisson}(\lambda T),
            \quad
            \mathbb{E}[N^i(T)] = \mathrm{Var}\bigl(N^i(T)\bigr) = \lambda T.
        \end{equation}
        \noindent
        The probability that $S^i$ never faces an audit in $[0,T]$ is $\exp(-\lambda T)$.\\
        \\
        Simply put, if an adversarial strategy deviates maliciously immediately after its last audit, the probability it remains unchecked for time $T$ is: $P_{\mathrm{undetected}}(T) = \exp(-\lambda T)$. This means the expected first time to audit is $1/\lambda$, and raising $\lambda$ increases the adversary’s expected cost of evasion (since each audit may incur dispute‐resolution fees) \footnote{Here we can trigger audits by the host blockchain network's VRF, ensuring that neither proposers nor verifiers can predict or bias audit selection.},\footnote{We can also make it such that only verifiers with minimum staking history or reputation score $\rho_v\ge\rho_{\min}$ may collect audit lotteries.}.
    \end{proposition}

    \noindent
    We pre-fund an $\mathcal L\subset\mathbb T$ lottery pool with native tokens; each audit draws a fixed on-chain fee $\phi_{\rm gas}$ plus a reward $\beta_{\rm al}\,A^i$ to participating verifiers, calibrated so that $\mathbb{E}[\phi_{\rm gas} + \beta_{\rm al} A^i]\ll A^i$.
    
\section{Alpha Token Dynamics}
    OpenAlpha employs two distinct tokens to decouple native-chain economics from protocol governance and verification:

    \begin{enumerate}
        \item Host‐Chain Token (SUPRA): The host blockchain network’s native token, SUPRA, serves as the economic spine for on-chain execution and strategy proposers’ skin-in-the-game:
        \begin{itemize}
            \item Collateral Staking: To submit a strategy, proposers must lock a predetermined amount of SUPRA, thereby deterring spam and frivolous submissions by imposing real economic risk.
            \item Proposer Commission: Upon delivering verifiably positive net value, a successful proposer earns a commission paid in SUPRA, compensating for deployment and transaction costs.
            \item Gas: All interactions (market trades, audits, governance votes) incur the usual host blockchain network denominated gas fees.
        \end{itemize}

        \item Reputation Token (Validation Rating Token): It is OpenAlpha protocol’s native utility and governance instrument, which must be held\footnote{Do we need to follow a decay schedule to ensure reputation that is accrued, after some reasonable time, is actually used and not just held?} or staked as a prerequisite for all model updates, parameter changes, and verification participation:
        \begin{itemize}
            \item Fees: Each prediction‐market transaction carries a validation fee, which funds protocol operations. Furthermore, voters in the prediction market who correctly vote, in turn earn newly minted Alpha tokens which can be seen as a form of rebate provided in AMMs to LPs.
            \item Validation Commission: A portion of SUPRA-based treasury commissions from strategy proposers are channeled to be staked into Alpha tokens. In this token, arbitrators are elected by the Alpha token holders, as they have the most incentives for the system to operate correctly. If the arbitrator layer is compromised no capital owner will use the protocol leading to full depreciation of the token.
            \item Governance: Critical parameter updates are conducted by Alpha token weighted votes, such that voters must lock Alpha tokens for the proposal duration, aligning long-term stakeholder incentives with protocol stability.
            \item Verifier Rewards: Community verifiers and deep-search auditors earn Alpha tokens from the fee pool, proportional to their prediction accuracy and on-chain reputation, which formalises and monetises adversarial scrutiny.
            \item Intention Creation: Capital owners must pay a non-refundable in Alpha tokens (burn reputation tokens), and as an expression of seriousness, must deposit the initial amount of assets to be managed through the strategies in this framework, in some form.
        \end{itemize}
    \end{enumerate}

\section{OpenAlpha: Beyond Strategy Validation}
    Beyond its primary role in validating capital strategies, the OpenAlpha framework generalises to a protocol for collaborative, adversarially-validated logical development. Any agent can propose formal logic (ranging from strategy functions, allocation rules, protocol modules, to risk scoring algorithms) as a candidate object of community validation. These proposals are subjected to structured verification, incentive-aligned review, and, where applicable, on-chain implementation.

\subsection{Collaborative Logic Development}
    OpenAlpha's validation architecture supports collaborative development of dApps and allocation logic by enabling modular, adversarially-audited innovation. A participant may, for instance, propose a novel allocation function - extending classical portfolio theory to account for dynamic on-chain risk factors such as protocol slippage, validator liveness, or liquidity fragmentation. This allocation logic, formally defined (e.g., as a mapping $f: (\omega^i, \xi^i) \mapsto A^i$ over strategy confidence scores $\omega^i$ and environmental variables $\xi^i$), is published on-chain as a candidate meta-allocation rule. The proposer stakes reputation-weighted tokens to instantiate a binary validation market, where mathematically-inclined peers assess the function’s performance claims—typically by verifying outperformance over benchmark portfolios, stress-tested robustness, or compatibility with declared intention sets. This market mechanism yields a confidence score $\Omega(f)$, representing the crowd-aggregated belief in the function's suitability for deployment.\\
    \\
    Subsequently, a separate participant, say someone specialised in smart contract engineering, may implement the validated function as an executable module, submitting it for adversarial verification. Deep searchers and community auditors then formally evaluate the implementation’s correctness, gas efficiency, and invariance to adversarial inputs (e.g., under MEV conditions or transaction reordering). Discrepancies or violations are financially incentivised to be identified and resolved via challenge-response games, with resolution outcomes feeding back into the function’s composite score. Over successive iterations, this process enables decentralised refinement of allocation logic across theoretical, empirical, and implementation layers. The result is a dynamically evolving library of rigorously validated allocation mechanisms, shaped by distributed expertise and governed by adversarial consensus, constituting a live, modular research substrate for on-chain collaboration.\\
    \\
    While OpenAlpha is designed around capital strategy validation, its architecture generalises to a domain-agnostic, cooperatively-validated adversarial framework for modular logic development. Any formalisable framework (whether a portfolio heuristic, smart contract, consensus algorithm, or application subcomponent) can be proposed, linked to a functional intention, and subjected to economic validation. This enables decentralised workflows where multiple agents contribute to different stages of development: one authoring mathematical logic, another implementing it, a third performing formal verification or empirical testing, and a fourth auditing for correctness or adversarial robustness. Each contribution is economically incentivised, publicly attestable, and independently validated, with the system coordinating partial work across contributors while preserving coherent verification guarantees. In effect, OpenAlpha functions as a distributed protocol for collaborative software and mechanism design, where validation itself is a programmable primitive and correctness emerges through structured, incentive-aligned peer review.

\bibliographystyle{plain}
\bibliography{main}

\begin{thebibliography}{1}

\bibitem{hanson2003combinatorial}
Robin Hanson.
\newblock Combinatorial information market design.
\newblock {\em Information Systems Frontiers}, 5:107--119, 2003.

\bibitem{koessler2008parimutuel}
Fr{\'e}d{\'e}ric Koessler, Charles Noussair, and Anthony Ziegelmeyer.
\newblock Parimutuel betting under asymmetric information.
\newblock {\em Journal of mathematical Economics}, 44(7-8):733--744, 2008.

\end{thebibliography}

\end{document}